\documentstyle[foga]{article} 

\input{psfig.sty}
\input epsf

\def\s{\scriptstyle}
\def\ss{\scriptscriptstyle}

\newcommand{\be}{\begin{eqnarray}}
\newcommand{\ee}{\end{eqnarray}}
\newcommand{\nn}{\nonumber}

\def\slashxi{{\xi}\!\!\!/}

\def\o{\over}
\def\C{{\s C}}

\def\px{P(\xi,t)}
\def\pxl{P(\xi,t+1)}

\def\pxf{P'(\xi,t)}

\def\pxc{P_c(\xi,t)}
\def\pnxc{P_c(\slashxi_i,t)}

\def\pxLf{P'(\xi_{\ss L},t)}
\def\pxRf{P'(\xi_{\ss R},t)}

\def\fav{{\bar f}(t)}
\def\fxi{{\bar f}(\xi,t)}
\def\frat{{\fxi\over \fav}}
\def\fx{{f_{\xi}}}
\def\fxL{{f_{\xi_L}}}
\def\fxR{{f_{\xi_R}}}

\def\dfx{\delta\fx}
\def\dfxL{\delta\fxL}
\def\dfxR{\delta\fxR}

\def\mutis{{\cal P}({ \xi})}
\def\mutijs{{\cal P}({ \slashxi_i\ra\xi})}

\def\ra{\rightarrow} 
 
\def\ef{f_{\ss\rm eff}(\xi,t)}
 
 \def\efff{f'_{\ss\rm eff}(\xi,t)}
\def\pxLft{P'(\xi_{\ss L},t')}
\def\pxRft{P'(\xi_{\ss R},t')}
\def\pxLft{P'(\xi_{\ss L},t')}
\def\pxRft{P'(\xi_{\ss R},t')}

\title{ Schemata as Building Blocks: Does Size Matter?} 
 
\author{ {\bf C. R. Stephens}\thanks{\ \ e-mail: stephens@nuclecu.unam.mx}
\ {\rm and} {\bf H. Waelbroeck}\thanks{\ \ e-mail: hwael@nuclecu.unam.mx} \\  
Instituto de Ciencias Nucleares, UNAM\\   
Circuito Exterior, A.Postal 70-543 \\ 
M\'exico D.F. 04510 \\ 
{\bf R. Aguirre}\thanks{\ \ e-mail: rosalia@nuclecu.unam.mx}\\
DEPFI, UNAM, A. Postal 70-543\\
M\'exico D.F. 04510\\}
 
\begin{document} 
 
\maketitle 

\begin{abstract} 

We analyze the schema theorem and the building block hypothesis using
a recently derived, exact schemata evolution equation.
We derive a new schema theorem based on the concept of 
{\it effective fitness} showing that schemata
of higher than average effective fitness receive an exponentially increasing 
number of trials over time. The building block hypothesis is a natural
consequence in that the equation shows how fit schemata are constructed
from fit sub-schemata. However, we show that generically there is no 
preference for short, low-order schemata. In the case where
schema reconstruction is favoured over schema destruction
large schemata tend to be favoured. As a corollary of the evolution equation
we prove Geiringer's theorem. We give supporting numerical evidence for
our claims in both non-epsitatic and epistatic landscapes.

\end{abstract} 
%\centerline{Keywords: Schema theorem, building blocks, effective fitness, 
%Geiringer's theorem.}
 
\section{Introduction} 

A very large proportion of scientific endeavour has been 
associated with the question: What are ``things'' made of? the reason
being that this is an indispensable requirement for understanding how
and why a ``thing'' functions. 
The answer has always tended to be: ``things'' are made of other, more 
elementary ``things''. In the physical sciences this is obvious:
a table is made up of atoms, which in turn are made up of electrons and
a nucleus, which in its turn ... In biology a living organism (generally) is 
composed of various organs and tissues, which in their turn are made 
of various types of cells, which in their turn are composed of various 
constituents such as nuclei, protoplasm, mitochondria etc. From the
nucleus we pass to chromosomes, genes, DNA and RNA etc. The
latter, along with other important ingredients, forming an elaborate
``computer programme'' for the construction of the organism. In computer
science high level languages are composed of more elementary 
languages until we arrive at the most basic machine level language 
recognized by the computer itself. 

What have all these examples in common? They show that all things
are made out of ``building blocks'', whether they be tables, giraffes or
computer programmes. Inevitably there exists a hierarchy
of building blocks, the hierarchy being ordered more often than not according
to scale and complexity. One can think of building blocks as the 
``effective degrees of freedom'' (EDOF) of a system, which in their turn are
composed of more ``fine grained'', elementary degrees of freedom. 
For complex systems the former are composed of very large numbers
of the latter. What building blocks one uses to describe a system 
depends very much on what one wants to say about it. 
In particular, on how fine grained a description one requires. 
Almost always a more coarse grained description will suffice. 

So what has the above to do with genetic algorithms (GAs)? 
At the most basic level all the above can be coded as bit 
strings (of course) and in some way or other be associated 
with the notion of adaptation and optimization of some ``fitness'' 
function in a complicated environment. Trying to understand these problems in
adaptation and optimization at the level of the fundamental, microscopic degrees
of freedom is prohibitively difficult: there are simply too many and, 
more often than not, they interact in a
highly non-linear fashion. In order to describe these systems  
both qualitatively and quantitatively 
one needs to know how EDOF emerge. In particular, in the context of GAs, 
if we coded the above problems as such how would some of the known EDOF
emerge? In GAs one can universally represent EDOF as
schemata. Of course, not all are of equal utility. EDOF if
they are to be useful in the description of a system must display a
certain degree of integrity. 

The question is: what schemata are utilized by a GA? Or rather, 
what are the typical properties possessed by successful schemata.
Naturally, the answer to this question will depend on the fitness
landscape of interest. However, one might enquire as to if or not
such properties exist in generic classes of landscapes. 
In fact theory has tried to be even more ambitious. The schema 
theorem and the related building block hypothesis \cite{holland}, \cite{goldberg}
propose that the EDOF in GAs are fit, {\it short} schemata irrespective
of the landscape! This is an extremely strong statement. The fact
that fit schemata are preferred is intuitively understandable, though
we will see some counterexamples to this later, whilst the purported
preference for short schemata is a supposed consequence of 
the destructive effect of crossover.

In this paper we will investigate theoretically and experimentally the
evolution of schemata and in particular how this evolution depends
on the defining length of the schemata. Our theoretical analysis will
be based on an exact evolution equation \cite{stewael1,stewael2,alt} 
for schemata for the case of proportional selection and $1$-point
crossover. In section 2 we will give a brief overview of the equation.
In section 3 we will discuss some of its more important theoretical 
ramifications and in section 4 we will show how numerical experiments
confirm the theoretical predictions. Finally in section 5 we will draw 
some conclusions.

\section{ Schema Equation} 

In this section we give without proof (see \cite{stewael1}, \cite{stewael2} 
for more details) the schema evolution equation for the relative proportion,
$P(\xi,t)\equiv n(\xi,t)/n$, of the schema $\xi$ of defining length $l$ and order $N_2$ 
at time $t$ in a canonical GA of population $n$ consisting of chromosomes
of $N$ bits evolving with respect to proportional selection, point mutation and $1$-point
crossover. In the limit $n\ra\infty$ $P(\xi,t)$ gives the probability of finding
the schema $\xi$ at time $t$. Explicitly
\be
\pxl=\mutis\pxc + \sum_{\s \slashxi_i}\mutijs\pnxc\label{maseqtwo}
\ee
where the sum is over all schemata, $\slashxi_i$, that differ
by at least one bit in one of the defining bits of the schema $\xi$,
and the effective mutation coefficients $\mutis$ and $\mutijs$ 
represent the probabilities that the schema $\xi$ remains unmutated
and the probability that the schema $\slashxi_i$ mutates to the schema $\xi$
respectively. $P_c(\xi,t)$ is the mean proportion of schemata $\xi$ at time $t$
after selection and crossover. Explicitly 
\be
\pxc= \pxf - {p_c\o N-1}\sum_{k=1}^{l-1}
\left(\pxf-P'(\xi_l(k),t)P'(\xi_R(l-k)\right)\label{eqth}
\ee
where $\pxf=(\fxi/\fav)\px$, $\fxi$ being the mean fitness of the schema $\xi$ and 
$\fav$ the average population fitness. $p_c$ is the crossover probability 
and $k$ the crossover point. The quantities $\pxLf$ and $\pxRf$ are defined 
analogously to $\pxf$ but refer to the schemata $\xi_L$ and $\xi_R$ which are 
the parts of $\xi$ to the left and right of $k$ respectively. 
One can illustrate the content of the equation with a simple diagramatic
example: $****1**|0**1*****$ is a schema with $l=7$ and $N_2=3$. The
crossover point is at $k=7$ hence $\xi_L$ has $N_2=1$ and $l=1$ while
$\xi_R$ has $N_2=2$ and $l=4$.
Note that the equation takes into account exactly both the effects 
of schema destruction and schema reconstruction. At the level of strings the 
equation will be equivalent to other exact formulations 
\cite{gold87,whitley,goldbridge,vose1,vose2,vose3}, and is most closely related 
to the analogous equation for a canonical GA evolving with respect to 
proportional selection and recombination derived by Altenberg \cite{alt} 
based on earlier work in genetics by Karlin and Liberman
\cite{karlin}.

There are several notable
features of the above equation: first of all it implies that crossover as an operator
imposes the idea of a schema. This can easily be seen by considering the above
equation for the case where $\xi$ is the entire string, $\C_i$. The reconstruction 
probability depends on the relative fitness of strings that contain the constituent 
elements, $\C_i^L$ and $\C_i^R$ of $\C_i$, but given that there can be many 
strings that contain $\C_i^L$ or $\C_i^R$ one must take an average over these 
strings. In this sense we are averaging over the ``degrees of  freedom'' represented 
by the bits that are not contained in $\C_i^L$ or $\C_i^R$. The equation also 
shows that the effects of reconstruction will 
outweigh destruction if the parts of a string are more selected than the whole.
This is closely related to the notion of linkage disequilibrium from population
genetics. However, linkage disequilibrium there is measured by the 
covariance, $C(\xi_L,\xi_R)\equiv P(\xi,t)-P(\xi_L,t)P(\xi_R,t)$, of the 
relative frequencies of $\xi_L$ and $\xi_R$. Here, we see that the relevant 
measure of schemata growth is the covariance, 
$C'(\xi_L,\xi_R)\equiv P'(\xi,t)-P'(\xi_L,t)P'(\xi_R,t)$, of the fitness-weighted
relative frequencies of $\xi_L$ and $\xi_R$. Clearly $C'$ does not have to be of the
same sign as $C$. 

The equation also shows an hierarchy in structure both 
in complexity and size and also in time. 
This is obvious in the very nature of the equation which relates a schema $\xi$ to
its ``building blocks'' of lower order and smaller defining length, $\xi_L$ and $\xi_R$. 
These in turn are related to even smaller, lower order building blocks $\xi_{LL}$, 
$\xi_{LR}$, $\xi_{RL}$ and $\xi_{RR}$ which are associated with reconstruction of the
schemata $\xi_L$ and $\xi_R$. The hierarchy terminates at $1$-schemata, i.e. 
schemata  with $N_2=1$, which are immune to the effects of crossover. 
The hierarchical nature of the evolution in terms of complexity and size 
is manifest in (\ref{eqth}) as schemata of a certain ``size''' (defining length and 
order) are related to schemata of  smaller size, which in their turn are related to 
yet smaller schemata etc. The structure is also temporal as schemata at time $t$ 
are related to smaller schemata at time $t-1$, which in turn are related to yet smaller 
schemata at time $t-2$ etc. 

Note the ``form invariance'' of the equation under a coarse 
graining. What does this mean? Consider the equation at the level of a complete string.
As we have pointed out, the very notion of recombination introduces a coarse graining
in that we can write the recombination contribution in terms of $P(\C_i^L,t)$ and 
$P(\C_i^R,t)$, which involve summing over the microscopic (bit) degrees of 
freedom of $\{\C_i-\C_i^L\}$ and $\{\C_i-\C_i^R\}$ respectively, where $\{-\}$ denotes set
difference. Coarse graining here simply means that we have forfeited detailed knowledge 
about the microscopic degrees of freedom not contained in $\C_i^L$ or $\C_i^R$. 
The corresponding evolution equations for $\C_i^L$ and $\C_i^R$ involve
recombination terms where a further coarse graining must be carried out 
via a summation, for example, over the microscopic
degrees of freedom of $\{\C_i-\C_i^{LL}\}$ in the case of the schema $\C_i^{LL}$. 
However, irrespective of the degree of coarse graining the form of the evolution
equation remains exactly the same. 

We see then that a general form of the building block hypothesis is inherent in the
very structure of the evolution equation for a canonical GA. Recombination 
builds complex schemata from more primitive consituents which in turn are 
constructed from yet more elementary building blocks until we arrive at the 
ultimate building blocks --- $1$-schemata. The question of whether a GA utilizes
building blocks to find a good solution can be seen to be related to whether
schema reconstruction or destruction is the most important effect. However, 
this is clearly not the only criterion. We have said that the evolution equation 
contains a generalized form of the block hypothesis: simply that larger, more
complex schemata are constructed from more primitive building blocks irrespective of
whether they are fit schemata utilized by the GA in finding fit chromosomes. This is 
not however the only way to grow a schemata. Perhaps the schemata was in
the initial population and was of high fitness. To see whether indeed a GA uses fit, short
building blocks as the standard building block hypothesis and Schema theorem purport
we must examine more closely the idea of fitness.

\section{What do we mean by fit?} 

Why are certain schemata preferred over others? Because they are fitter
of course. But what does one really mean by this statement? Consider the
following contrived but instructive example: 
consider a $2$-schemata, i.e. $N_2=2$, problem with crossover 
but neglecting mutation, and with a fitness landscape where $f(01)=f(10)=0$ and
$f(11)=f(00)=1$. The steady state solution of the schema evolution equation is
\be
P(11)=P(00)={1\over2}\left(1-{p_c\over2}{(l-1)\over(N-1)}\right)\quad\quad
P(01)=P(10)={p_c\over4}{(l-1)\over(N-1)}
\ee
For $l=N$ and $p_c=1$ we see that half the steady state population is composed
of strings that have zero fitness! Note that this fixed point is in fact a 
stable one. Such results lead one to doubt whether the
concept of fitness is the most relevant one in gauging the growth of a schema. 

Another, more relevant, example is associated with a GA of binary 
alleles with mutation and proportional selection but without crossover 
(the Eigen model \cite{eigen}). Consider a ``needle-in-a- 
haystack'' type fitness landscape where there exists one string, $\C_m$ 
--- the ``master sequence'', 
of high fitness all the rest being of equal low fitness. When the mutation rate, $\mu$, 
is zero then the steady state population, assuming $n$ is large, is such 
that $P(\C_m)=1$. When $\mu>0$ but small, 
$P(\C_m,t\ra\infty)\neq 1$. However, the population is clustered around the master 
sequence in that the Hamming distance between the master sequence and the large 
majority of other strings in the steady state population is small. Increasing $\mu$ one
reaches a critical value, $\mu_c$, beyond which $P(\C_m,t\ra\infty)\ra 1/2^N$, which
is exactly the proportion expected in a completely random population. The sharp 
phase transition at $\mu_c$ is familiar from thermodynamics being due to the 
competition between ``energy'' (selection) and ``entropy'' (mutation). In fact, 
the quantity $-(1/2)\ln(\mu/(1-\mu))$ is the precise analog of the thermodynamic 
temperature. For $\mu\geq\mu_c$
the evolution of the GA cannot be well understood by thinking of evolution on the 
needle-in-a-haystack landscape. Thermodynamically this corresponds to considering
the energy landscape as opposed to the more physically relevant free energy landscape. 
In the entropy dominated regime for $\mu\geq\mu_c$ every state has the same free energy
hence there is effectively no selection acting. Once again we are led to call into question
the usefulness of the standard notion of fitness. 

As a third simple example consider the effect of mutation without 
crossover in the context of
a model that consists of $2$-schemata, $11$, $01$, $10$, $00$, where
each schema can mutate to the two adjacent ones when the states
$11$, $10$, $00$, $01$ are placed clockwise on a circle. 
For example, $11$ can mutate to $10$ or $01$ but not to $00$. 
We assume a simple degenerate fitness landscape: $f(11)=f(01)=f(10)=2$, $f(00)=1$.
Clearly there is no selective advantage for any one of the three
degenerate schemata over the others.
In a random population, $P(11) = ... = P(00) = {1 \over 4}$. 
If there is uniform probability, $\mu$, for each schema to mutate to an adjacent
one then the evolution equation that describes this system is
\be
P(i,t+1)=(1-2\mu)P'(i,t)+\mu(P'({i-1},t)+P'({i+1},t))
\ee
For $\mu=0$ the steady state population is $P(11)=P(01)=P(10)=1/3$, 
$P(00)=0$. Thus we see the synonym symmetry of the 
landscape associated with the degeneracy of the states $11$, $10$ and $01$
is unbroken. However, for $\mu>0$, the schema distribution at $t=1$ 
starting from a random distribution at $t=0$ is $P(11)=2/7$, 
$P(01)=P(10)=(2-\mu)/7$, $P(00)=(1+2\mu)/7$.
Thus, we see that there is an induced breaking of the landscape synonym 
symmetry due to the effects of mutation. In other words the population 
is induced to flow along what in the fitness landscape is a flat direction. 

Why is it that the examples above lead us to reconsider the idea of fitness? Fitness is
intrinsically associated in the standard picture with a particular 
genetic operator --- selection.
In the above we have two examples that exhibit regimes where other genetic operators, 
crossover and mutation respectively, can dominate and another example wherein
populations are forced to flow along directions with zero selection gradient. 
In such regimes intuition gleaned from
the normal fitness landscape is of little value. Clearly what is 
required is a generalization
of fitness, a type of ``effective fitness'', that treats the various 
genetic operators on a more democratic footing and where population 
flows take place in an ``effective fitness'' 
landscape. In the case of selection and mutation the problem can be reformulated 
into a problem in equilibrium thermodynamics \cite{leut} hence the standard thermodynamic 
free energy may be utilized. In more general circumstances one must find a more general
effective fitness. 

A natural candidate for an effective fitness has been given in
\cite{stewael1,stewael2,stewael3}. Specifically, 
\be
\pxl={\ef\o\fav}\px\label{three}
\ee
Comparing with equation (\ref{maseqtwo}) one finds
\begin{eqnarray}
\ef=\mutis\left({P_c(\xi,t)\over P(\xi,t)}\right)\fxi
+\sum_{\s \slashxi_i}\mutijs
\left({P_c(\xi,t)\over P(\xi,t)}\right)\fav
\label{effit}   
\end{eqnarray}
This definition is very natural from an evolutionary viewpoint as it 
gives a direct measure of the reproductive success of a given schema.
In the limit $\mu\ra0$, $p_c\ra0$ one finds that $\ef\ra\fxi$. 

Using the concept of effective fitness one can much better understand the 
three examples given earlier. For instance, in the first example, although
the schemata $01$ and $10$ have zero fitness their effective fitness is
non-zero. Similarly, in the third example although the schemata $11$, $10$ 
and $01$ are degenerate in terms of fitness this degeneracy is lifted
by the effect of mutation and this is manifest at the level of the
effective fitness. The above also leads to the idea of an effective selection 
coefficient, $s_{\ss\rm eff}=\ef/\fav-1$, that measures directly selective
pressure including the effects of genetic operators other than selection. 
If we think of $s_{\ss\rm eff}$ 
as being approximately constant in the vicinity of time $t_{\ss 0}$, 
then $s_{\ss\rm eff}(t_0)$ gives us the exponential rate of increase or 
decrease of growth of the schema $\xi$ at time $t_0$.
In the limit of a continuous time evolution the solution of 
(\ref{three}) is
\be
P(\xi,t)=P(\xi,0){\rm e}^{\int_0^ts_{\ss\rm eff}dt'}\label{four}
\ee

Using the evolution equation and the concept of effective fitness one
can formulate a new Schema theorem that unlike the standard schema theorem
is an equality rather than an inequality 
\vskip 0.3cm
\noindent{\bf Schema Theorem}
\be
\pxl={\ef\o\fav}\px\label{schtheorem}
\ee
\vskip 0.3cm
The interpretation of this equation is clear and analogous to the 
old schema theorem: schemata of higher 
than average {\it effective fitness} will be allocated an ``exponentially'' 
increasing number of trials over time. We put the word exponentially
in quotes as the real exponent, $\int^t s_{\ss{\rm eff}}dt'$, is not, 
except for very simple cases such as a flat fitness landscape,
of the form $\alpha t$, where $\alpha$ is a constant. The above
says much more than the standard schema theorem: first of all
it is an exact equation not just a lower bound; secondly it gives
a deeper insight into the role of crossover. This comes about because
the equation takes into account schema creation. The standard schema
theorem emphasizes only the destructive effect of crossover. 

Armed with the above we can much more readily investigate the reconstructive aspect
of crossover. In fact we will see generically that it is a more important 
effect. This is of great relevance in the link between the schema 
theorem and the building block hypothesis. The former gives a quantitative
bite to the latter by showing that the destructive effects of crossover are
greater the longer the defining length of the schema; thus leading to the notion that
{\it short}, fit schemata are favoured. We can readily see that this is not
generally true. If schema reconstruction outweighs that of destruction then
crossover is more positive the longer the schema. Under such circumstances
{\it long}, fit schemata are favoured. We will see this confirmed experimentally
in the next section. Finally, we may remark that unless the fitness landscape
warrants it there is no reason to think of a building block as a ``local'' object,
as seems to be the case in the work on Royal Road functions \cite{mitch} 
where attempts were made to validate the building block hypothesis by giving 
high fitness to a very small number of states associated with localized blocks 
of $1$s.

Another natural definition of effective fitness which leads us to a very simple proof
of Geiringer's theorem \cite{geir} follows from splitting the
evolution equation into those terms that are linear
in $P(\xi,t)$ and those that are independent of it.
For instance, in the case of selection and crossover we have 
\be
\pxl= {\efff\o\fav}\px+j(t)\label{effeq}
\ee
where $\efff = (1- p_c{(l-1)\o N-1}){\fxi\o\fav}$ and 
$j(t)={p_c\o N-1}\sum_{k=1}^{l-1}\pxLf\pxRf$. The corresponding
effective selection coefficient is 
$s'_{\ss{\rm eff}} = ((1- p_c{(l-1)\o N-1}){\fxi\o\fav}-1)$. In the limit
of a continuous time evolution (\ref{effeq}) may be formally integrated to yield
\be
P(\xi,t)={\rm e}^{\int_0^t s'_{\ss{\rm eff}}(t')dt'}P(\xi,0)+
{\rm e}^{\int_0^t s'_{\ss{\rm eff}}(t')dt'}
\int^t_0j(t'){\rm e}^{-\int_0^{t'} s'_{\ss{\rm eff}}(t'')dt''}dt'
\label{formsol}
\ee
In a flat landscape $s'_{\ss{\rm eff}}=-p_c(l-1)/(N-1)$, hence
\begin{eqnarray}
P(\xi,t)={\rm e}^{-p_c{(l-1)\o(N-1)}t}P(\xi,0)+
{p_c\over (N-1)}{\rm e}^{-p_c{(l-1)\o(N-1)}t}\times\nn \\
\sum_{k=1}^{l-1}\int^t_0\pxLft\pxRft
{\rm e}^{p_c{(l-1)\o(N-1)}t'}dt'
\end{eqnarray}
Notice that dependence on the initial condition, $P(\xi,0)$, is exponentially
damped unless $\xi$ happens to be a $1$-schema, the solution of the 
$1$-schemata equation being $P(i,t)=P(i,0)$.
An immediate consequence is that when considering the source term describing
reconstruction the only non-zero terms that need to be taken into account are
those which arise from $1$-schemata, as any higher order term will always
have an accompanying exponential damping factor. Thus, we see that the fixed
point distribution for a GA with crossover evolving in a flat fitness landscape
is
\be
P^*(\xi)=Lt_{t\ra\infty}P(\xi,t)=\prod_{i=1}^{N_2}P(i,0)
\ee
which is basically Geiringer's Theorem in the context of schema
distributions and simple crossover. We see here that the theorem appears in 
an extremely simple way as a consequence of the solution of the evolution equation.
Note that the fixed point distribution arises purely due to the effects of 
recombination, the long time behaviour depending only on the initial distribution
of the most elementary building blocks --- the $1$-schemata.

Geiringer's theorem will also hold in a non-flat landscape if selection is 
only very weak, where what we mean by weak is that $\frat\sim(1+\epsilon)$ and 
$\epsilon<(p_c((l-1)/ (N-1))/ 1-p_c((l-1)/ (N-1)))\ \  \forall l>1$. 
In this case anything other than a $1$-schema will once again be 
exponentially damped. In this case however due to the non-trivial landscape certain 
$1$-schemata are preferred over others. A concrete example of such a landscape is
$f_i=1+\alpha_i$ where $\sum_i\vert\alpha_i\vert\leq(\epsilon/(2+\epsilon))$ 
and $f_i$ is the fitness of the ith bit. 
Note there is no need to restrict to a linear fitness function here, arbitrary epistasis
is allowed as long as it does not lead to large fitness deviations away from the mean.
In this case $\frat<1+\epsilon$. 

So what can we glean from our evolution equation in terms of schema size vis
a vis the building block hypothesis? For a flat fitness landscape the effective 
selection coefficient for a $2$-schema $ij$ is
\be
s_{\ss\rm eff}=-p_c\left({l-1\o N-1}\right)+p_c\left({l-1\o N-1}\right)
{P(i,0)P(j,0)\o P(ij,t)}
\ee
One sees that schema reconstruction will exceed that of schema destruction
if $i$ and $j$ are negatively correlated, i.e. if the linkage disequilibrium
is negative. When will this be the case? The
fitness landscape itself can quite easily induce negative correlations between
$\xi_L$ and $\xi_R$. In this case there is a competition 
between the (anti-) correlating effect of the landscape and 
the mixing effect of crossover. For instance, in the neutral case of a 
Kaufmann $k=0$ landscape 
when $\dfxL, \dfxR > 0$, where $\dfxL=\fxi-\fav$, 
then $1 + {2N_2 \o N} \dfx < (1 + {2N_L \o N} \dfxL)
(1 + {2N_R \o N} \dfxR)$, so selection induces an anti-correlation, hence 
in an uncorrelated initial population, 
$P'(\xi, t) < P'(\xi_L, t) P'(\xi_R,t)$. In this case there will clearly be a 
preference for large schemata rather than small ones. In the case of deceptive
problems, for instance the minimal deceptive problem \cite{gold87}, typically
there is a positive correlation. Under such circumstances schema destruction 
will be the dominant effect and one should see a preference for short, fit 
schemata as hypothesized by the canonical Schema theorem and Block Hypothesis.
In a fitness landscape that has both a deceptive and a non-deceptive
component a given schema may typically be reached via deceptive or non-deceptive
channels. For instance, for a $3$-schema, $ijk$, perhaps the channel $ij+k\ra ijk$
is deceptive whereas the channel $i+jk\ra ijk$ is not. If both channels
are equally likely in terms of selection then crossover will favour the 
non-deceptive channel and this will be manifest in the fact that one
channel contributes positively to the effective fitness whereas the other contributes 
negatively. It has been hypothesized \cite{stewael3} that generically
populations will evolve via non-deceptive channels. Further 
theoretical results for $k=0,2$ can be found in \cite{stewael1,stewael2}.

\section{Experimental Results} 

In this section we present experimental evidence for many of the statements
and theoretical results we have presented. Given our desire to 
investigate the effects of crossover vis a vis its effect on schema length we
considered a GA with $\mu=0$ as mutation, being a local operator, can 
have no direct effect on schema length. We will first consider the case of
a non-epistatic landscape --- the well known counting ones, or unitation problem. We 
considered a population of $5000$ $8$-bit strings, thus the maximum 
schema length is $8$. We chose a large population so as to be able to ignore
finite size effects. Figures 1 and 2 are plots of $M(l)$ versus time where
$M(l)\equiv (n_{\ss opt}(l)-n_{\ss opt}(8))/n_{\ss opt}(8)$. Here, 
$n_{\ss opt}(l)$ is the number of optimal $2$-schemata of defining length $l$ 
normalized by the total number of length $l$ 2-schemata per string, i.e. $9-l$. 
By optimal $2$-schemata we mean schemata containing the global optimum $11$.
$n_{\ss opt}(8)$ is the number of optimal $2$-schemata of defining length $8$.
Figure 1 is with $p_c=0$ and Figure 2 with $p_c=1$. We show averages over
$30$ different runs. Without crossover there is essentially no preference for 
schemata of a given length. Adding in crossover leads to a remarkable change: Schemata 
prevalence is ordered monotonically with respect to length but with the larger schemata
being favoured. This is in complete accord with the theoretical prediction of the
previous section based on considerations of the evolution equation.
We emphasize that this is purely an effect of the crossover operator and
shows clearly that schema reconstruction is more important than schema 
destruction. 

\begin{figure}[h]
$$
\psfig{figure=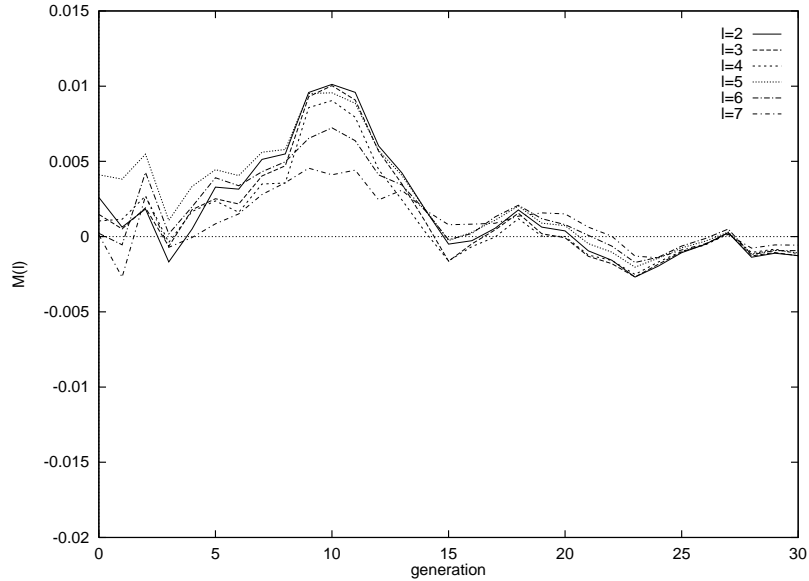,angle=-90,width=4.3in}
$$
\caption{Graph of $M(l)$ versus $t$ in the unitation model with $p_c=0$.}
\end{figure}
%\psfig{figure=f1a.ps}
%\centerline{Figure 1a: Graph of $M(l)$ versus $t$ in}
%\centerline{the unitation model with $p_c=0$. }
%{\epsfxsize=20cm

\begin{figure}[h]
$$
\psfig{figure=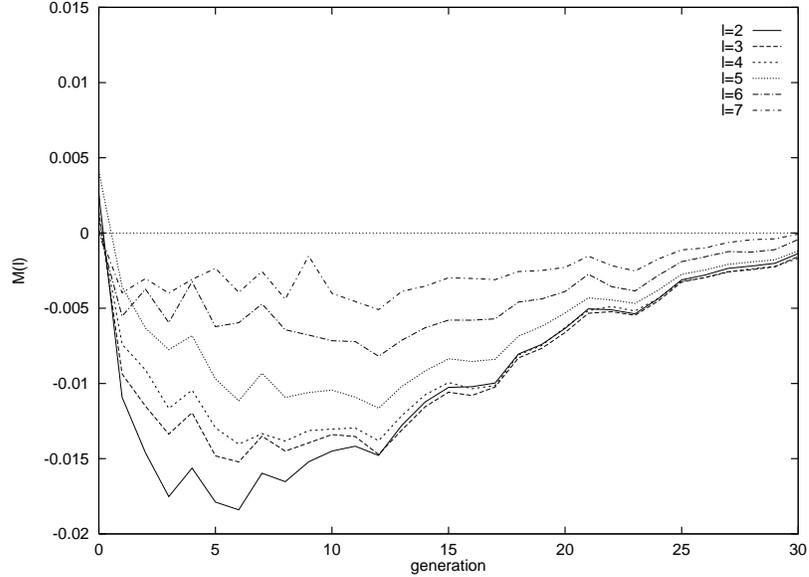,angle=-90,width=4.3in}
$$
\caption{Graph of $M(l)$ versus $t$ in unitation model with $p_c=1$.}
\end{figure}
%\epsfbox{f1b.ps}
%\psfig{figure=f1b.ps,width=4.3in}
%\centerline{Figure 1b: Graph of $M(l)$ versus $t$ in unitation model with $p_c=1$.}
%\centerline{the counting ones model with $p_c=1$. }

This effect is measured nicely by the effective fitness function in
Figures 3 and 4. What we in fact graph is 
$F(l)\equiv(f_{\ss\rm  eff}(l)-f_{\ss\rm  eff}(8))/f_{\ss\rm  eff}(8)$. 
where $f_{\ss\rm  eff}(l)$ is the effective fitness of optimal $2$-schemata
of size $l$ and $f_{\ss\rm  eff}(8)$ is the analogous quantity for 
optimal $2$-schemata of size $8$. 
Note that the effective fitness of larger schemata is greater than 
that of shorter ones for the first $6$ or so generations, in fact significantly so 
given that the fluctuations in Figure 3 are of the order of $0.0005$ whereas the 
relative effective fitness advantage of schemata of size $8$ relative to size $2$ is,
from Figure 4, about $0.01$, i.e. $20$ times larger! As can be seen from 
(\ref{four}) a positive selection coefficient is associated with a schema that is
growing in number relative to another. After $6$ generations the curves in Figure 2
start to converge again which coincides with the effective fitness now being larger for 
the smaller schemata. Roughly speaking one can think of the effective fitness as
being a measure of the gradient of the curves in Figures 1 and 2. If one repeats the 
experiments for schemata of order $3$ or higher one will once again find a preference
for long, ``non-local'' schemata; non-local in the sense that there is no preference
whatsoever for the three bits to be found together. 

\begin{figure}[h]
$$
\psfig{figure=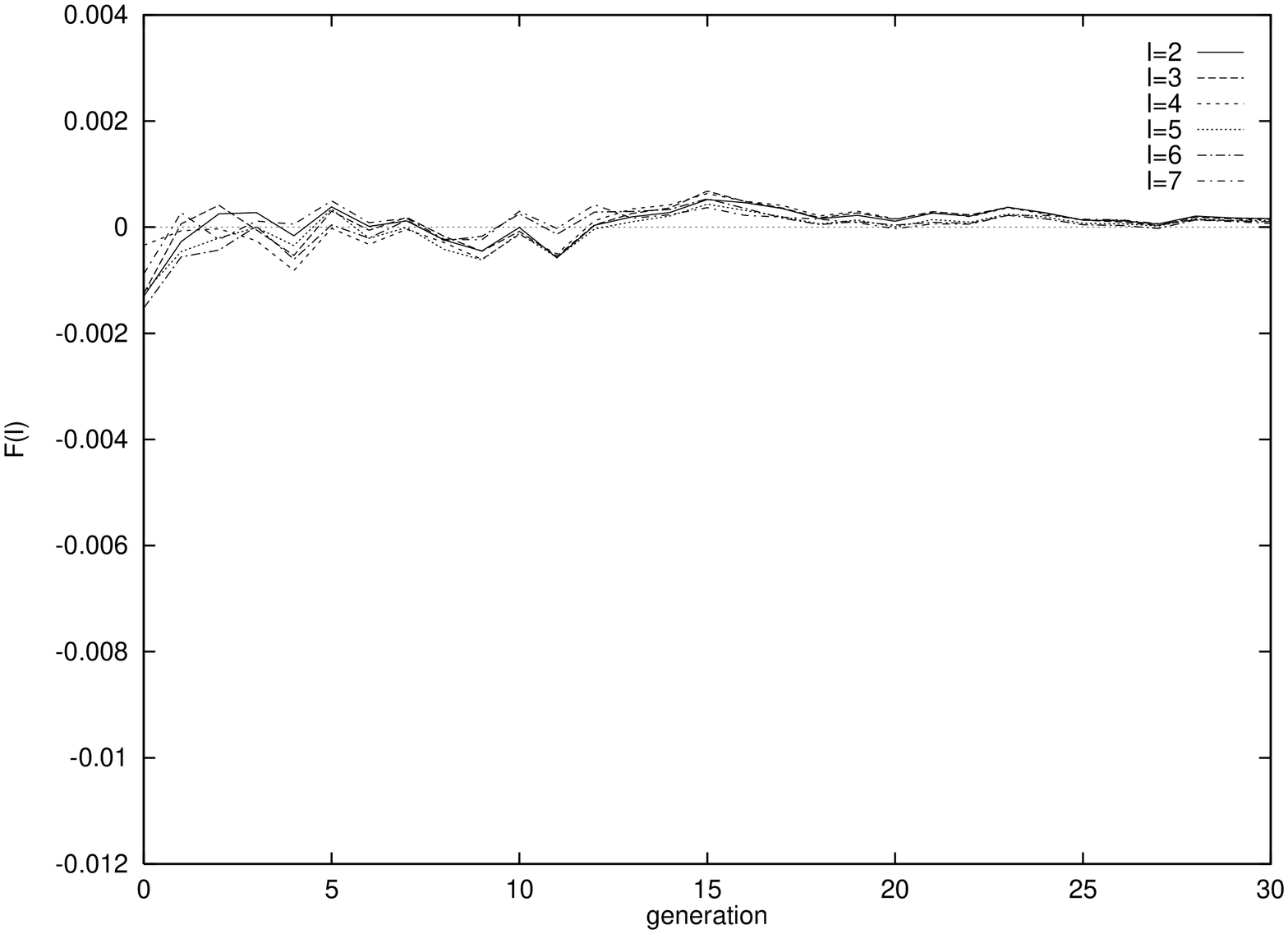,angle=-90,width=4.3in}
$$
\caption{Graph of $F(l)$ versus $t$ for unitation model with $p_c=0$.}
\end{figure}
%\epsfbox{f2a.ps}
%\psfig{figure=fig2a.ps,width=4.3in}
%\centerline{Figure 2a: Graph of $F(l)$ versus $t$ for unitation model with $p_c=0$.}
% optimal $2$-schemata as a function} 
%\centerline{of time in the counting ones model with $p_c=0$. }
%\vspace{2mm}

\begin{figure}[h]
$$
\psfig{figure=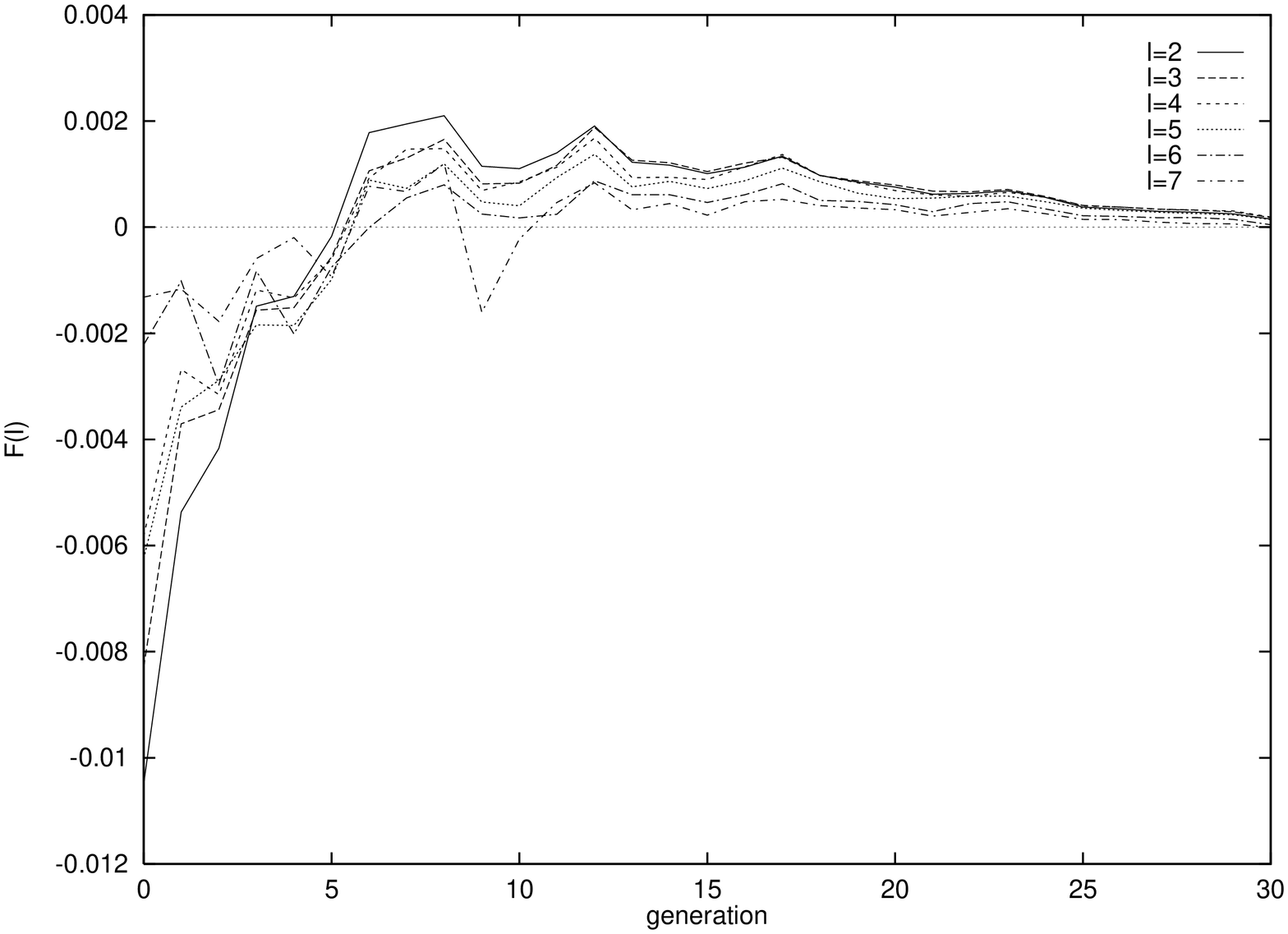,angle=-90,width=4.3in}
$$
\caption{Graph of $F(l)$ versus $t$ for unitation model with $p_c=1$.}
\end{figure}
%\psfig{figure=fig2b.ps,width=4.3in}
%\centerline{Figure 2b: Shows the effective fitness of schemata at each}
%\centerline{generation in the counting ones model of Figure 1b.}
%\newpage\
One might say that
this is all fine and good but we have only shown what happens for non-epistatic
landscapes. An important aspect of the counting
ones landscape is not so much that it is non-epistatic as rather it is neutral
in the absence of crossover, in that there is no preference in the landscape
itself for schemata of a certain length. This means we can study directly the
geometrical effect of crossover without having to worry about the complicating 
effects of selection. Having established generic properties of crossover we
can then turn to various classes of fitness landscape to investigate the
intricate relationship between the two operators.
We will first introduce epistasis by considering what happens in the case 
of landscapes of the form
\be
f(\C_i)=\sum_j1_j+{\epsilon\o N^{\pm}}\sum_{\ss jk\in\C_i}l_{jk}^{\pm1}
\ee
where the first sum is over all the $1$s of the string $\C_i$ and the second
is over all pairs of $1$s, $l_{jk}$ being the defining length of the schema
with $1$s at the points $j$ and $k$. $N^{\pm}=\sum_{jk}l_{jk}^{\pm1}$ is a 
normalization constant, being a sum over all optimal $2$-schemata of the form $11$
of the optimum string $11111111$. $\epsilon$ simply controls the size of the 
length-dependent epistatic term relative to the counting ones term. 
Summing over the lengths of the schemata gives a landscape where there
is a preference for long schemata. On the contrary, summing over the 
inverses of the lengths gives a landscape where there is a bias for short
schemata. In the former case the epistasis can be thought of as giving rise to an
effective repulsion between pairs of $1$s and in the latter an effective
attraction. In both cases the epistasis between string bits depends on their
distance apart.

The results can be seen in Figures 5-8. In Figure 5
we see the effect of a bias for large schemata of magnitude $\epsilon=0.3$ 
with $p_c=0$. The graph is quite similar 
to that of Figure 2 hence one can see that crossover leads to an effective 
bias for larger schemata that is similar in many respects to an effective 
repulsion between schema bits. In Figure 6 we see what happens when one
includes a bias for small schemata of magnitude $\epsilon=0.75$ with $p_c=1$. 
Note that the effect of crossover is to completely annul the effect of the 
landscape bias. In Figures 7 and 8 we see the evolution of $F(l)$
in both cases. In Figures 9 and 10 we see what happens in a deceptive landscape. 
The landscape we chose was one where for each of the 28 different pairs
of the $8$-bit string we have two possible sets of conditions on the fitness of
each pair: i) $f(11)=3$, $f(01)=f(10)=1$, $f(00)=2$; and ii)   
$f(11)=3$, $f(01)=f(10)=2$, $f(00)=1$. The first set clearly is deceptive,
the second clearly not. As a function of the total number of deceptive
pairs, $n_d$, we can vary how deceptive the total landscape is. For $n_d=28$
the landscape is totally deceptive, whilst for $n_d=0$ there is no 
deception. In Figure 10 we see the 
effects of crossover on a totally deceptive landscape. Note that the 
effect of crossover in this case is to increase the bias for short schemata
due to the fact that schema destruction is more important than schema
reconstruction.

\begin{figure}[h]
$$
\psfig{figure=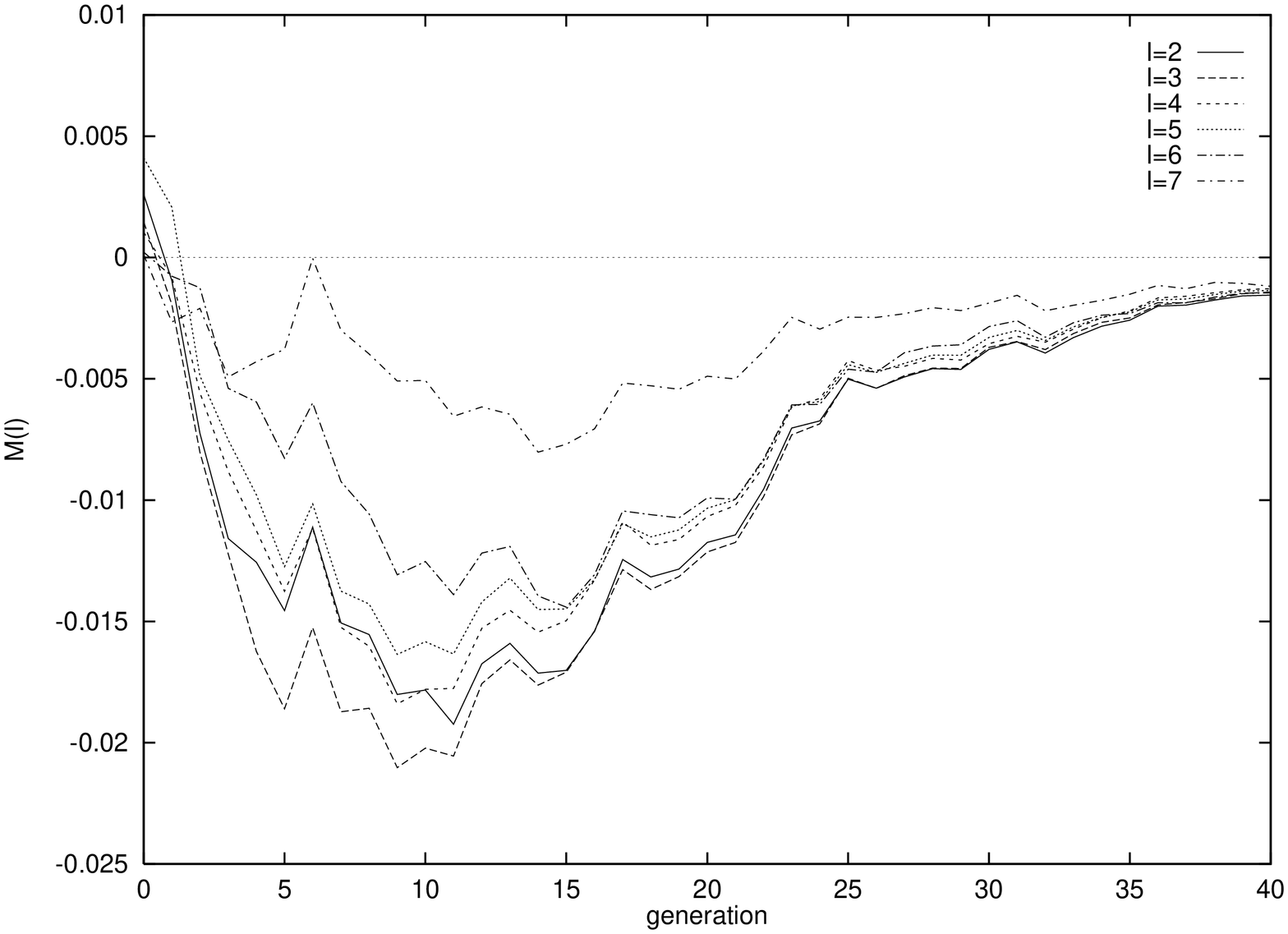,angle=-90,width=4.3in}
$$
\caption{Graph of $M(l)$ versus $t$ for biased model with bit-bit repulsion with 
$\epsilon=0.3$, $p_c=0$. }
\end{figure} 
%\psfig{figure=fig3a.ps,width=4.3in}
%\centerline{Figure 3a: Shows the evolution of schemata at each 
%generation in the biased model} 
%\centerline{where the longest schemata have the highest fitness. 
%In this case $\epsilon$ $=0.3$ and $pc=0.0$. }

\begin{figure}[h]
$$
\psfig{figure=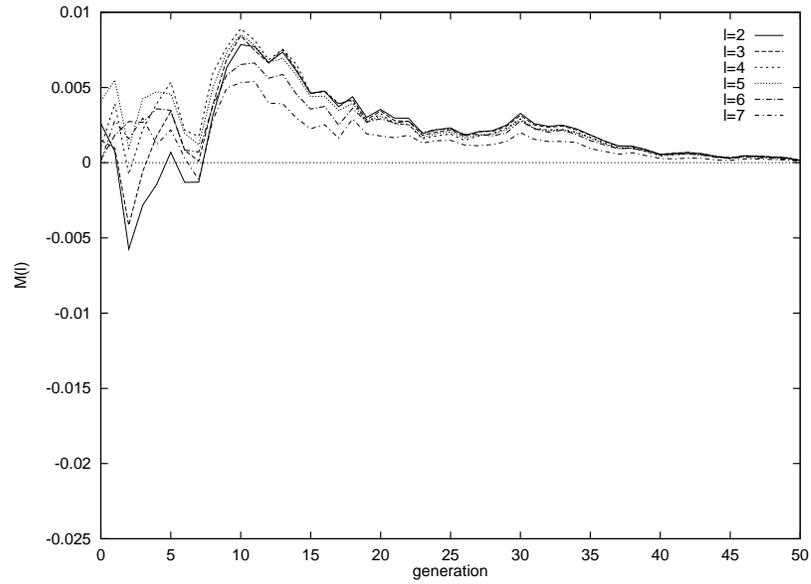,angle=-90,width=4.3in}
$$
\caption{Graph of $M(l)$ versus $t$ for biased model with bit-bit attraction with 
$\epsilon=0.75$, $p_c=1$.}
\end{figure}
%\psfig{figure=fig3b.ps,width=4.3in}
%\centerline{Figure 3b: Shows the evolution of schemata at each 
%generation in the biased model}
%$\centerline{ where the shortest schemata have the highest fitness. 
%$In this case $\epsilon$ $=0.75$ and $pc=1.0$. }

\begin{figure}[h]
$$
\psfig{figure=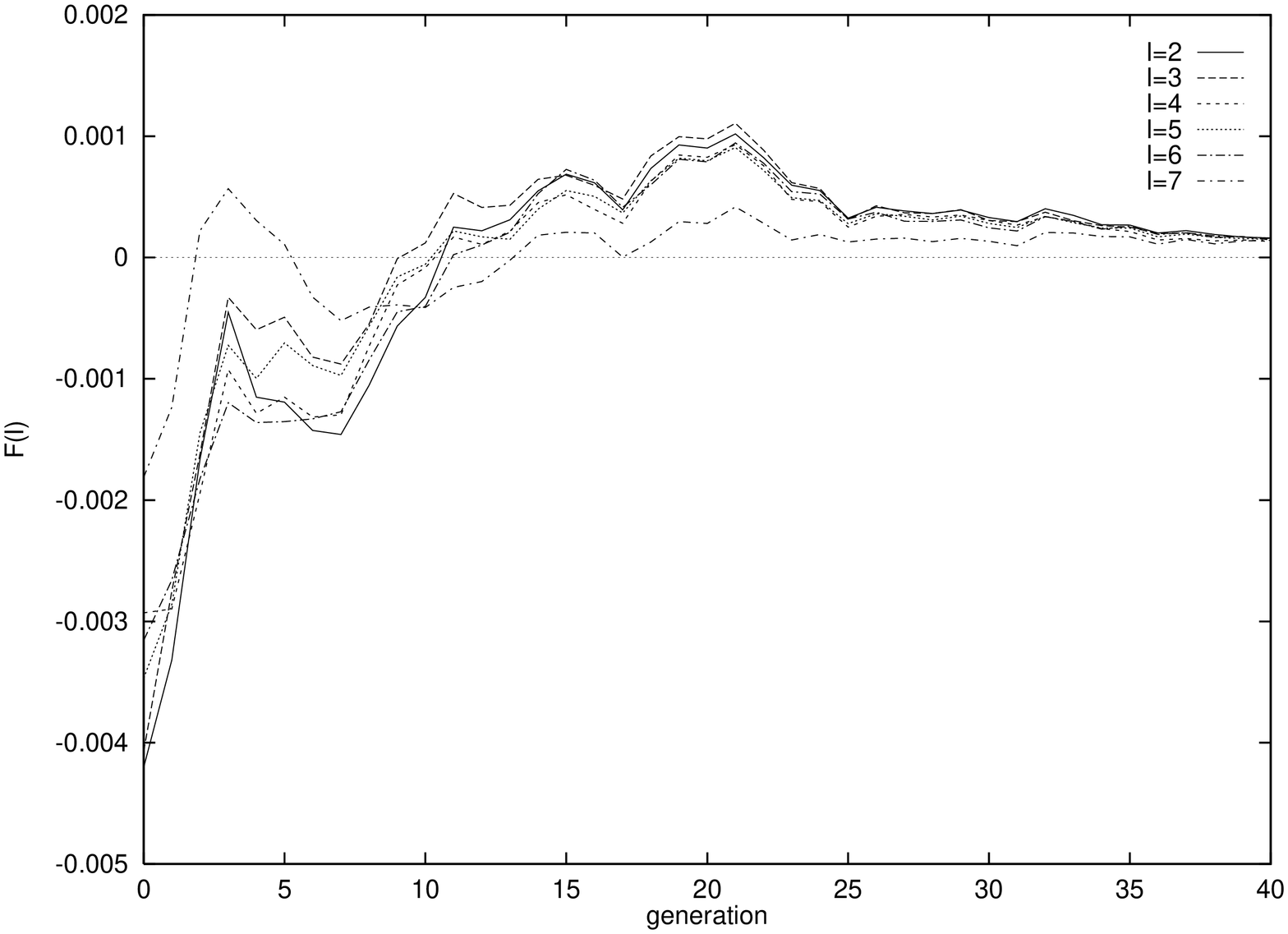,angle=-90,width=4.3in}
$$
\caption{Graph of $F(l)$ versus $t$ for biased model of Figure 5. }
\end{figure}
%\psfig{figure=fig4a.ps,width=4.3in}
%\centerline{Figure 4a: Shows the effective fitness of schemata at each}
%\centerline{generation in the biased model of figure 3a.}

\begin{figure}[h]
$$
\psfig{figure=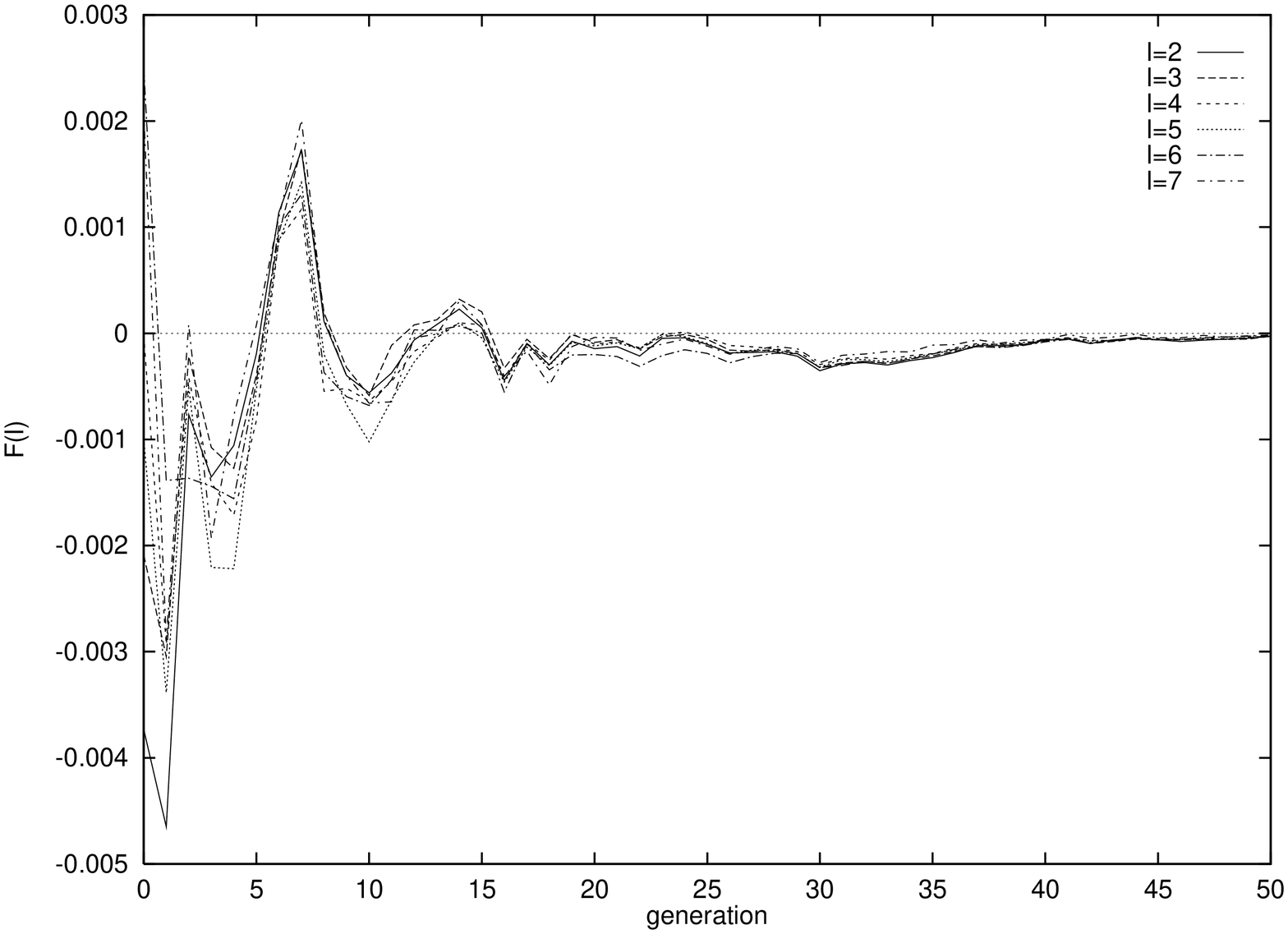,angle=-90,width=4.3in}
$$
\caption{Graph of $F(l)$ versus $t$ for biased model of Figure 6.}
\end{figure}
%\psfig{figure=fig4b.ps,width=4.3in}
%\centerline{Figure 4b: Shows the effective fitness of schemata at each}
%\centerline{generation in the biased model of figure 3b.}

\begin{figure}[h]
$$
\psfig{figure=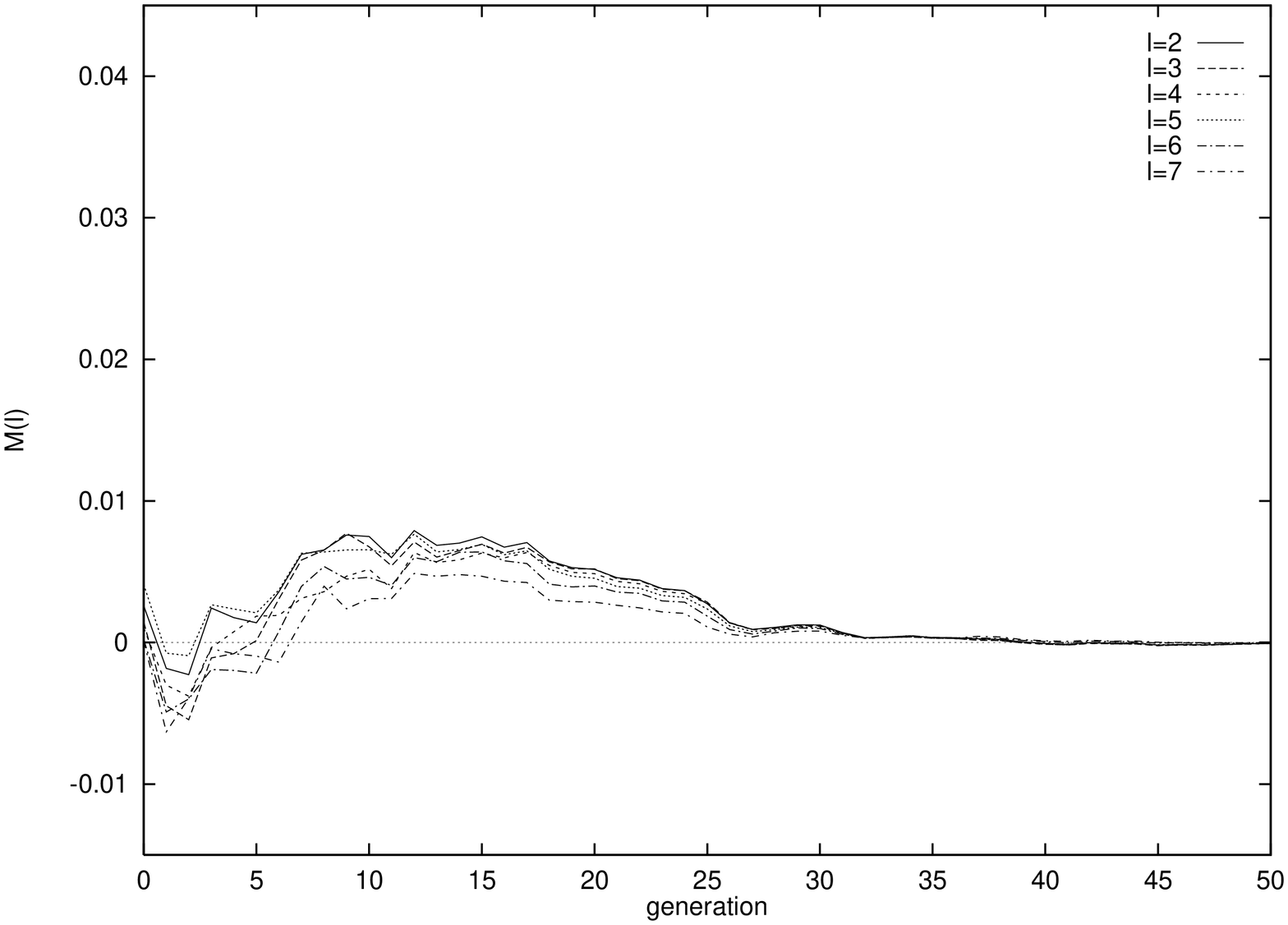,angle=-90,width=4.3in}
$$
\caption{Graph of $M(l)$ versus $t$ in the fully deceptive model with $p_c=0$.}
\end{figure}
%\psfig{figure=fig5a.ps,width=4.3in}
%\centerline{Figure 5a: Shows the evolution of schemata at each generation} 
%\centerline{in the deceptive model with crossover probability $pc = 0.0 $. }

\begin{figure}[h]
$$
\psfig{figure=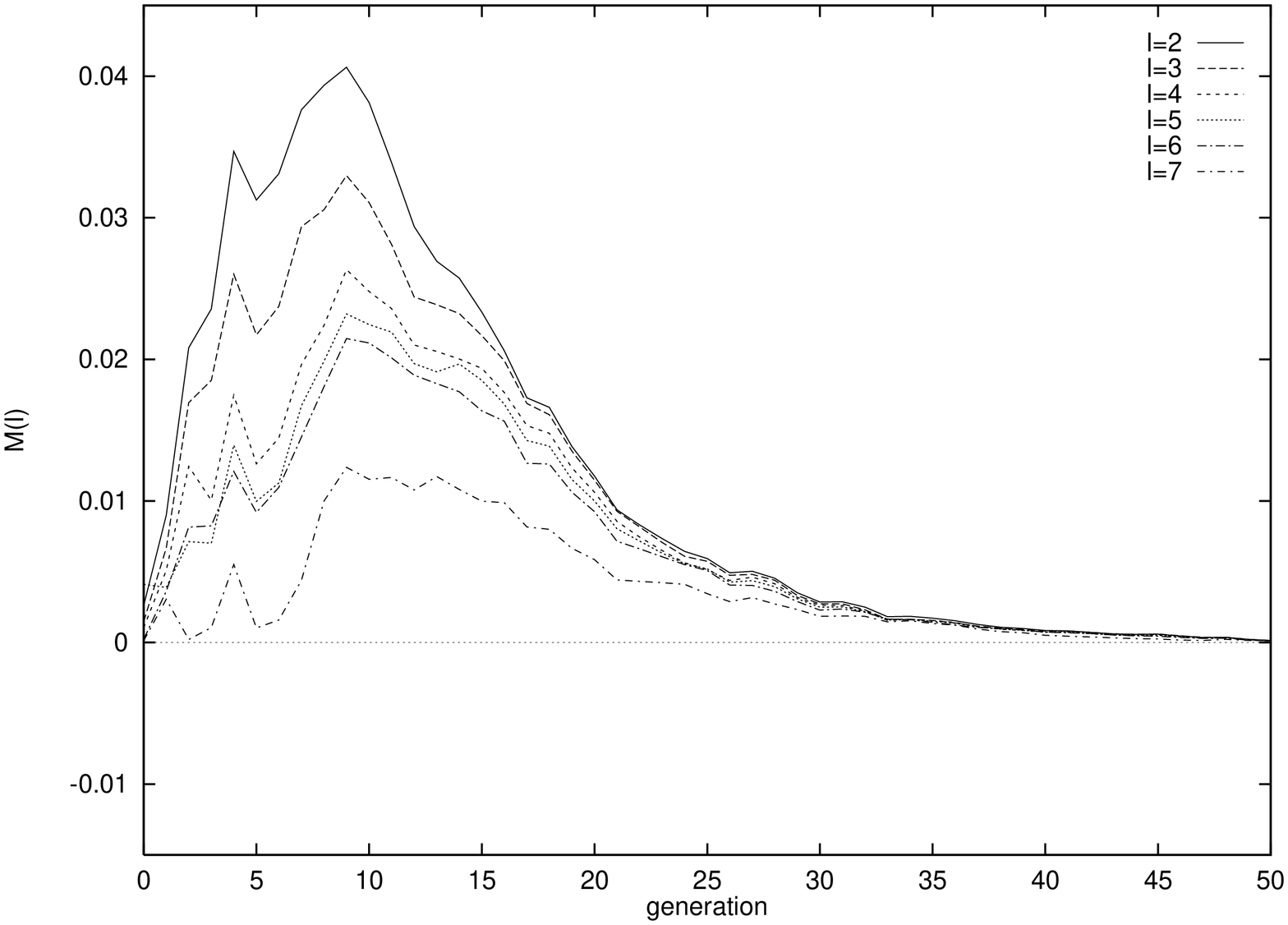,angle=-90,width=4.3in}
$$
\caption{Graph of $M(l)$ versus $t$ in the fully deceptive model with $p_c=1$.}
\end{figure}
%\psfig{figure=fig5b.eps,width=4.3in}
%\centerline{Figure 5b: Shows the evolution of schemata at each generation}
%\centerline{in the deceptive model with crossover probability $pc = 1.0 $. }

%\clearpage

\section{Conclusions}

In this paper we have analyzed the consequences of an exact evolution
equation for GAs which applies, in a very natural way, directly to schemata 
thus allowing for a critical analysis of the Schema 
theorem and the Building Block hypothesis. We saw that the very
structure of the equation, taking into account as it does schema 
reconstruction, contains a general form of the building block hypothesis; longer, 
higher order schemata being constructed from smaller, lower order schemata
when schema reconstruction dominates. The ultimate building blocks were
shown to be $1$-schemata as they are immune to the effects of crossover.

We noted that a building block interpretation was natural in the case where 
schema reconstruction dominates, irrespective of whether the blocks were fit
or not. In order to investigate under what conditions fit, small schemata 
were combined into fit, large schemata we found it useful to critically
examine the concept of fitness. We used explicit examples to demonstrate
that selective fitness and the corresponding fitness
landscape were inadequate to intuitively understand GA evolution.
We therefore introduced the notion of effective fitness, showing  
that it was a more relevant concept than pure selective fitness
in governing the reproductive success of a schema. Based on this concept
and 
%\clearpage
the evolution equation we introduced a new 
schema theorem that showed that schemata of high effective fitness 
receive an exponentially increasing number of trials as a function of time. 
We also showed that generically there is no preference for short, low-order 
schemata. In fact, if schema reconstruction dominates the opposite is true. 
Only in deceptive problems does it
seem that short schemata will be favoured, and then only in totally deceptive
problems, as the system will tend to seek out the non-deceptive channels
if they exist. 

We performed various experiments to verify our theoretical results in both
epistatic and non-epistatic landscapes. For non-epistatic landscapes
we confirmed that there is indeed a preference for large schemata. In fact
we showed that schema prevalence is a monotonically increasing function of
schema defining length. In a class of epistatic landscapes designed to 
give an effective repulsion or attraction between pairs of bits we showed
that crossover in its action was analogous to a bit-bit repulsion, thus
favouring long schemata. For a model deceptive landscape we showed that in the 
case of total deception contrary to all the previous cases, and as predicted
on theoretical grounds, short schemata were favoured.

It would naturally be very interesting to see
if other exact results besides Geiringer's theorem follow very simply from 
the evolution equation. A pressing matter is the search for approximation
schemes within which the equations can be solved, as for a general landscape
\clearpage
an exact solution will be impossible. In this respect techniques familiar
from statistical mechanics, such as the renormalization group might 
well prove very useful. Of course, much more 
experimental analysis is needed on a wider set of test landsacpes. 
In particular it will be of interest to
test the hypothesis that GAs will seek out non-deceptive trajectories if
possible. 
%\clearpage
\subsubsection*{Acknowledgements} 
This work was partially supported through DGAPA-UNAM grant number IN105197.
CRS is grateful to an anonymous referee for bringing the work
of Lee Altenberg to his attention and to Adam Pr\"ughel-Bennett for useful
comments on the manuscript.

\subsubsection*{} 

\vfill\eject
\end{document}